\documentclass[a4paper,11pt]{article}

\usepackage{jheppub}
\usepackage{graphicx,subfigure}
\usepackage[colorlinks=true,linktocpage=true,linkcolor=blue,citecolor=blue]{hyperref}

\def\be{\begin{eqnarray}}
\def\ee{\end{eqnarray}}

\def\nn{\nonumber\\}

\title{Quark Number Susceptibilities from Two-Loop Hard Thermal Loop Perturbation Theory}

\author[a]{Najmul Haque,}
\author[a]{Munshi G. Mustafa,}
\author[b]{Michael Strickland}

\affiliation[a]{Theory Division, Saha Institute of Nuclear Physics, Kolkata, India - 700064}
\affiliation[b]{Physics Department, Kent State University, Kent, OH 44242 United States}

\abstract{We use the recently obtained two-loop hard thermal loop perturbation theory thermodynamics functions of a plasma of quarks and gluons to compute the diagonal second- and fourth-order quark number susceptibilities.  The two-loop hard thermal loop perturbation theory thermodynamic functions used are reliable in the limit that the ratio of the quark chemical potential to temperature is small.  Using this result, we are able to obtain (semi-)analytic expressions for the quark number susceptibilities at leading- and next-to-leading-order in hard thermal loop perturbation theory.  We compare the hard thermal loop perturbation theory results with perturbative quantum chromodynamics calculations, a Polyakov-loop Nambu-Jona-Lasinio model calculation, and lattice quantum chromodynamics results.}

\begin{document}

\maketitle
\flushbottom


\section{Introduction}
\label{intro}

Dynamical chiral symmetry breaking and confinement are two well-known fundamental features of quantum chromodynamics (QCD). Nowadays it is generally believed that at high temperature and/or baryonic density strongly interacting matter will undergo chiral symmetry restoration and deconfinement phase transitions to a state of matter called the quark-gluon plasma (QGP).  The study of such phase transitions has become a field of both theoretical and experimental interest since information about these phase transitions has the potential to provide a more fundamental understanding of QCD itself.  In recent years, considerable effort has been dedicated to the creation of a QGP in the laboratory.  The collider experiments currently dedicated to this search are the Relativistic Heavy Ion Collider (RHIC) at Brookhaven National Lab (BNL) and the Large Hadron Collider (LHC) at the  European Organization for Nuclear Research (CERN).  Future experiments are planned at the Facility for Antiproton and Ion Research (FAIR) at the Gesellschaft f\"ur Schwerionenforschung (GSI) facility.  In all cases experimentalists collide heavy ions which have been accelerated to relativistic speeds in order to create the conditions necessary for the creation of a short-lived quark-gluon plasma.

In the confined/chiral-symmetry-broken phase, quarks are confined inside hadrons which possess integer-valued baryon number.  In the deconfined/chiral-symmetry-restored phase, quarks which have fractional baryon number are free to propagate over larger distance scales.  This fundamental difference leads to different quark number fluctuations in the two phases~\cite{jeon}. In general, fluctuations of conserved quantities~\cite{forster}, such as baryon number, electric charge, strangeness, isospin charge, etc. are considered to be an important diagnostic tool for investigating the quark-hadron phase transition in relativistic heavy ion collisions~\cite{mclaren,kunihiro}.  The magnitude of quark number fluctuations can be determined by computing the quark number susceptibilities (QNS), which measure the response of the quark number density to an infinitesimal change in the quark chemical potential in the limit of zero chemical potential.  In addition, it was also argued recently that the QNS may be used to identify the position of the critical end point in the QCD phase diagram~\cite{krishna}.  The QNS have been extensively studied in the last two decades using a variety of approaches including perturbative QCD (pQCD)~\cite{mclaren,toimela,kapusta,vourinen}, lattice QCD (LQCD) simulations~\cite{wp,peter,peter1,hotqcd,alton,bazavov,bernard,cheng,gavai}, Nambu-Jona-Lasinio (NJL) models~\cite{kunihiro,kusaka,ratti}, Polyakov Loop-Nambu-Jona-Lasinio (PNJL) models~\cite{pnjl,sandeep,paramita,ratti}, hard-thermal/dense-loop (HTL/HDL) resummation techniques~\cite{munshi,munshi1,blaizot,Rebhan:2003wn,jiang,su}, rainbow-ladder and beyond-rainbow-ladder approximations of the Dyson-Schwinger equations~\cite{rainbow}, strong-coupling techniques~\cite{holographic}, functional renormalisation group techniques~\cite{renorm}, and quasiparticle models \cite{quasi}.

In view of the ongoing experimental and theoretical effort to understand the phase structure of QCD, the determination of QNS using a variety of approaches is important.  Although the study of QCD matter very close to the phase transition requires a nonperturbative description, it is also interesting to explore the behavior of QNS beyond leading order (LO)~\cite{munshi,munshi1,blaizot,jiang,su} within weak coupling expansions which employ state-of-the-art resummation techniques~\cite{braaten,pisarski,andersen1,andersen2,andersen3}. Unlike LQCD, the weak-coupling expansion can straightforwardly handle finite density and temperature.  As a result, such calculations can provide useful information about QNS.  Herein we will assume massless quarks.  For massless quark flavors the QNS are usually defined as
\begin{eqnarray}
 \chi_{n}(T)\equiv \left. \frac{\partial^n\!{\cal P}}{\partial \mu^n} \right |_{\mu= 0} \ ,
\label{qns_def}
\end{eqnarray}
where $\cal P$ is the pressure of system, $\mu$ is the quark chemical potential and $T$ is the temperature of the system.  Above, $n$ is the order of the QNS and we note that all odd orders vanish at $\mu=0$ due to the charge-parity symmetry of the system.  The equation of state (EOS) expressed in terms of the pressure $\cal P$ of QCD matter at high temperature and density is an important quantity, which we have recently computed to next-to-leading order (NLO)~\cite{haque} at nonzero $\mu$ and $T$ employing the hard thermal loop perturbation theory (HTLpt) reorganization of finite temperature QCD~\cite{braaten,pisarski,andersen1,andersen2,andersen3}. In this paper we compute the second- and fourth-order QNS from the LO and NLO HTLpt pressure and compare our results with conventional perturbative QCD calculations, various LQCD results, and a Polyakov-loop Nambu-Jona-Lasinio model.       

The paper is organised as follows: In Sec.~\ref{htlpt} we briefly review the HTLpt formalism. In Sec.~\ref{p_htlpt} we 
present (semi-)analytic expressions for the LO and NLO HTLpt pressure.  In Sec.~\ref{qns} the second- and fourth-order QNS at LO and NLO are derived. The corresponding results are discussed and compared with first principles LQCD calculations, conventional pQCD, and a PNJL model.  We conclude and give an outlook for future work in Sec.~\ref{concl}.


\section{Hard Thermal Loop Perturbation Theory}
\label{htlpt}

HTL perturbation theory~\cite{andersen1,andersen2,andersen3} is a reorganization of the perturbation
series for hot and dense QCD which has the following Lagrangian density
\begin{eqnarray}
{\cal L}= \left({\cal L}_{\rm QCD}
+ {\cal L}_{\rm HTL} \right) \Big|_{g \to \sqrt{\delta} g}
+ \Delta{\cal L}_{\rm HTL} \, ,
\label{L-HTLQCD}
\end{eqnarray}
where $\Delta{\cal L}_{\rm HTL}$ collects all necessary additional HTL renormalization counterterms and 
${\cal L}_{\rm HTL}$ is the HTL effective Lagrangian~\cite{braaten,pisarski}.  The HTL effective Lagrangian can be
written compactly as
\begin{eqnarray}
\label{L-HTL}
{\cal L}_{\rm HTL}=-{1\over2}(1-\delta)m_D^2 {\rm Tr}
\left(G_{\mu\alpha}\left\langle {y^{\alpha}y^{\beta}\over(y\cdot D)^2}
	\right\rangle_{\!\!y}G^{\mu}_{\;\;\beta}\right)
	+ (1-\delta)\,i m_q^2 \bar{\psi}\gamma^\mu \left\langle {y_\mu\over y\cdot D}
	\right\rangle_{\!\!y}\psi
	\, ,  \label{htl_lag}
\end{eqnarray}
where $D$ is the covariant derivative, $y^\mu=(1,{\mathbf y})$ is a light like vector and 
$\langle \cdots \rangle$ is the average over
all possible directions, ${\hat y}$, of the loop momenta.
The HTL effective action is gauge invariant, nonlocal, and can generate all 
HTL $n$-point functions~\cite{braaten,pisarski}, which satisfy the necessary 
Ward-Takahashi identities by construction. 
The mass parameters $m_D$ and $m_q$ are the Debye gluon screening and  quark masses 
in a hot and dense medium, respectively,  which depend on the strong coupling $g$, the temperature $T$, and 
the chemical potential $\mu$.   In the end we will formally treat the masses $m_D$ and $m_q$  
in (\ref{L-HTL}) as being leading-order in $g$ in order to make the calculation tractable~\cite{andersen1,andersen2,andersen3}.
The $n^{\rm th}$ loop order in the HTLpt loop expansion is obtained by expanding the partition function through order $\delta^{n-1}$
and then taking $\delta \rightarrow 1$~\cite{andersen1,andersen2,andersen3,munshi,munshi1,munshi2,jiang,su,haque}.
In higher order calculations,  one usually fixes the parameters $m_D$ and $m_q$ by employing a variational prescription which
requires that the first derivative of the pressure with respect to both $m_D$ and $m_q$ vanishes
such that the free energy is minimized~\cite{andersen1,andersen2,haque}.  In the following section, we briefly describe 
the recently computed finite temperature and chemical potential NLO HTLpt pressure~\cite{haque}.  


\section{Leading- and Next-to-leading-order HTLpt Pressure}
\label{p_htlpt}

Using the above reorganization of finite temperature/density QCD, the two-loop pressure for a plasma of quarks and gluons 
can be obtained in HTLpt by expanding in the ratio of chemical potential to temperature through fourth order
in $\mu/T$.
Below we quote both leading-order (LO) and next-to-leading-order (NLO) expressions for the pressure, along with the relevant mass gap equations, which are then 
used to compute NLO QNS at various orders.  The calculation of the LO and NLO HTLpt pressure at finite temperature and chemical potential was presented in an earlier paper \cite{haque} and we refer the reader to this paper for the detailed calculation.


\subsection{LO Pressure}

The LO HTLpt pressure through ${\cal O}(g^4)$ at any $\mu$ is~\cite{su,haque}
\be
{\cal P}_{\rm LO}& =&  d_A\frac{\pi^2T^4}{45}\left\{1+\frac{7}{4}\frac{d_F}{d_A}\left(
1+\frac{120}{7}\hat\mu^2+\frac{240}{7}\hat\mu^4\right) - \frac{15}{2} \hat m_D^2 
-30\frac{d_F}{d_A}\left(1+12\hat\mu^2\right)\hat m_q^2
\right.\nn&+&\left.
30\hat m_D^3+\frac{45}{4}\left(\ln{\frac{\hat\Lambda}{2}}-\frac{7}{2}+\gamma+\frac{\pi^2}{3}\right)
\hat m_D^4 - 60\frac{d_F}{d_A}(\pi^2-6)\hat m_q^4\right\}
 \;.
\label{pressure-LO}
\ee
Note that apart from the free contribution, no explicit terms proportional to $\hat\mu^4$ appear through ${\cal O}(\hat m_q^4)$.
The dimensionless variables $\hat m_D$, $\hat m_q$, $\hat\Lambda$, and $\hat \mu$ are defined as 
\be
\hat m_D &=& {m_D \over 2 \pi T}  \;,
\hspace*{0.2in}
\hat m_q = {m_q \over 2 \pi T}  \;,
\hspace*{0.2in}
\hat \Lambda = {\Lambda \over 2 \pi T}  \;,
\hspace*{0.2in}
\hat \mu = {\mu \over 2 \pi T}  \;. 
\ee
At leading order, the weak coupling expressions for the mass parameters are 
\be 
m_D^2 &=& \frac{g^2T^2}{3} \left [ c_A +s_F\left(1+12\hat\mu^2\right)  \right ]\, ;  \hspace*{0.2in}
m_q^2 = \frac{g^2T^2}{4} \frac{c_F}{2} \left (1 + 4\hat\mu^2\right ) \, . \label{mass_lo}
\ee
In the expressions for the pressure and masses above we use the standard notation for the various Casimir invariants necessary:
$d_F=N_cN_f$, \  $d_A=N_c^2-1$, \  $s_F=N_f/2$ , $c_A=N_c$ and $c_F=(N_c^2-1)/2N_c$.  At LO
the variational method for fixing the mass parameters $m_D$ and $m_q$ can not be used in practice, since it results
in only the trivial solution $m_D=m_q=0$.  Instead at LO one canonically uses the LO masses listed in Eqs.~(\ref{mass_lo}) as the lowest order ``variational solutions'' for the mass parameters.


\subsection{NLO HTLpt Pressure and Variational Mass Gap Equations}

The NLO HTLpt pressure through ${\cal O}[(\mu/T)^4]$ is~\cite{haque}
\be
{\cal P}_{\rm NLO}&=&
           d_A {\pi^2 T^4\over45} \Bigg\{ 
	   1 + {7\over4} {d_F \over d_A}\left(1+\frac{120}{7}
           \hat\mu^2+\frac{240}{7}\hat\mu^4\right) - 15 \hat m_D^3 
\nn && \hspace{2mm} - 
	   {45\over4}\left(\log\hat{\Lambda\over2}-{7\over2}+\gamma+{\pi^2\over3}\right)\hat m_D^4	
	   + 60 {d_F \over d_A}\left(\pi^2-6\right)\hat m_q^4
\nn && \hspace{1mm} +	
           {\alpha_s\over\pi} \Bigg[ -{5\over4}\left(c_A + {5\over2}s_F\left(1+\frac{72}{5}
           \ \hat\mu^2+\frac{144}{5}\ \hat\mu^4\right)\right) 
	   + 15 \left(c_A+s_F(1+12\hat\mu^2)\right)\hat m_D
\nn && \hspace{1.5cm} - 
	 {55\over4}\left\{ c_A\left(\log{\hat\Lambda \over 2}- {36\over11}\log\hat m_D - 2.001\right)
		- {4\over11} s_F \left[\left(\log{\hat\Lambda \over 2}-2.333\right)\right.\right.
\nn && \hspace{2.5cm} + 
	\left.\left.  (24-18\zeta(3))\left(\log{\hat\Lambda \over 2} -15.662\right)\hat\mu^2 \right. \right.
\nn && \hspace{3.5cm} + \left. \left.
	120\left(\zeta(5)-\zeta(3)\right)\left(\log{\hat\Lambda \over 2} -1.0811\right)\hat\mu^4 + 
         {\cal O}\left(\hat\mu^6\right)\right] \!\!\right\} \hat m_D^2
\nn && \hspace{2mm} - 
	45 \, s_F \left\{\log{\hat\Lambda\over 2} + 2.198 - 44.953\hat\mu^2 - \left(288 \ln{\frac{\hat\Lambda}{2}} 
       +19.836\right)\hat\mu^4 + {\cal O}\left(\hat\mu^6\right)\right\}\hat m_q^2
\nn && \hspace{2mm} +
	 {165\over2}\left\{ c_A\left(\log{\hat\Lambda \over 2}+{5\over22}+\gamma\right) \right.
\nn && \hspace{2mm} - \left.
     {4\over11} s_F \left(\log{\hat\Lambda \over 2}-{1\over2}+\gamma+2\ln2 -7\zeta(3)\hat\mu^2+
         31\zeta(5)\hat\mu^4 + {\cal O}\left(\hat\mu^6\right) \right)\right\}\hat m_D^3
\nn && \hspace{2mm} +
         30 s_F \left(\frac{\zeta'(-1)}{\zeta(-1)}
         +\ln \hat m_D\right) \! \! \left[(24-18\zeta(3))\hat\mu^2 + 120(\zeta(5)-\zeta(3))\hat\mu^4 + 
           {\cal O}\left(\hat\mu^6\right)\right] \hat m_D^3 \nonumber \\ 
&& \hspace{3mm}
	+ 180\,s_F\hat m_D \hat m_q^2 \Bigg]
\Bigg\} \; , 
\label{pressure-NLO}
\ee
which is accurate up to ${\cal O}(g^3)$ and nominally accurate to ${\cal O}(g^5)$ since it was obtained from an expansion of two-loop 
thermodynamic potential in a power series in $m_D/T$ and $m_q/T$ treating both $m_D$ and $m_q$ having leading terms proportional to $g$.
By ``nominally accurate'' we mean that we expand the scalar integrals treating $m_D$ and $m_q$
as ${\cal O}(g)$ keeping all terms which contribute through ${\cal O}(g^5)$; however, 
the resulting series is accurate to order $g^5$ in name
only.  At each order in HTLpt the result is an infinite series in $g$. Using the mass 
expansion we keep terms through order $g^5$ at all loop-orders of HTLpt in order to make
the calculation tractable.  At LO one obtains only the correct perturbative coefficients 
for the $g^0$ and $g^3$ terms when one expands in a strict power series in $g$.  At NLO 
one obtains the correct $g^0$, $g^2$, and $g^3$ coefficients and at NNLO one obtains
the correct $g^0$, $g^2$, $g^3$, $g^4$, and $g^5$ coefficients.  The resulting approximants
obtained when going from LO to NLO to NNLO are expected to show improved convergence since the loop
expansion is now explicitly expanded in terms of the relevant high-temperature degrees of
freedom (quark and gluon high-temperature quasiparticles).
  
Using the result above, the mass parameters $m_D$ and $m_q$ can be obtained using the variational prescription
\begin{eqnarray}
  \left.\frac{\partial}{\partial m_D}{\cal P}_{\rm NLO}\right|_{m_q=const.}=0 \ \ \ \ \ \mbox{and}\ \ \ \ \  
 \left.\frac{\partial}{\partial m_q}{\cal P}_{\rm NLO}\right|_{m_D=const.}=0 \; , 
\end{eqnarray}
which leads to two gap equations for $m_D$ and $m_q$, respectively
\begin{eqnarray}
 && \hspace{-7mm} 
 45\hat m_D^2\left[1+\left(\ln\frac{\hat\Lambda}{2}-\frac{7}{2}+\gamma+
\frac{\pi^2}{3}\right)\hat m_D\right]
\nonumber\\&=&
\frac{\alpha_s}{\pi}\Bigg\{15(c_A+s_F(1+12\hat\mu^2))
-\frac{55}{2}\left[c_A\left(\ln\frac{\hat\Lambda}{2}-\frac{36}{11}\ln{\hat m_D}-3.637\right)
\right.
\nonumber\\
&-&\left.\frac{4}{11}s_F\left\{\ln\frac{\hat\Lambda }{2}-2.333+(24-18\zeta(3))
\left(\ln\frac{\hat\Lambda }{2}-15.662\right)\hat\mu^2
\right.\right.
\nonumber\\
&+&\left.\left.120(\zeta(5)-\zeta(3))\left(\ln\frac{\hat\Lambda }{2}-1.0811\right)\hat\mu^4
\right\}\right]\hat m_D
+\frac{495}{2}\left[c_A\left(\ln\frac{\hat\Lambda }{2}
+\frac{5}{22}+\gamma\right)
\right.
\nonumber\\
&-&\left.
\frac{4}{11}s_F\left\{\ln\frac{\hat\Lambda }{2}-\frac{1}{2}
+\gamma+2\ln2-7\zeta(3)\hat\mu^2+31\zeta(5)\hat\mu^4
\right.\right.
\nonumber\\
&-&\left.\left.
\left(\frac{\zeta'(-1)}{\zeta(-1)}+\ln \hat m_D+
\frac{1}{3}\right)\!\left((24-18\zeta(3))\hat\mu ^2+120(\zeta(5)-\zeta(3))\hat\mu^4\right)\right\}\right] \hat m_D^2
+180 s_F\hat m_q^2\Bigg\}, \nonumber \\
\label{gap_md}
\end{eqnarray}
and
\begin{eqnarray}
 \hat m_q^2&=&\frac{d_A}{8d_F\left(\pi ^2-6\right)}\frac{\alpha_s s_F}{\pi }\left[3\left(\ln\frac{\hat\Lambda }{2}
+2.198-44.953\ \hat\mu^2 \right. \right.
\nonumber \\
&& \left. \left.
\hspace{4cm}
-\left(288\ln\frac{\hat\Lambda }{2}+19.836\right)\hat\mu^4\right)-12\hat m_D\right]. 
\nonumber \\
\label{gap_mq}
\end{eqnarray}

To obtain the second and fourth-order quark number susceptibilities in HTLpt, one requires 
expressions for  
$m_D$,  $\frac{\partial^2}{\partial\mu^2}m_D$,  $m_q$, and $\frac{\partial^2}{\partial\mu^2}m_q$ 
at $\mu=0$ from Eqs.~(\ref{gap_md}) and (\ref{gap_mq}).\footnote{Note that odd derivatives 
with respect to $\mu$ vanish at $\mu=0$. Fourth-order derivatives at $\mu=0$
are nonzero, however, they appear as multiplicative factors of the gap equations and are therefore not required, as we will see below.}
We list these here for completeness.  The result
for the limit of the $m_D$ gap equation necessary is
\begin{eqnarray}
 && \hspace{-1cm} 
 45\hat m_D^2(0)\left[1+\left(\ln\frac{\hat\Lambda}{2}-\frac{7}{2}+\gamma+
\frac{\pi^2}{3}\right)\hat m_D(0)\right]
\nonumber\\&=&
\frac{\alpha_s}{\pi}\Bigg\{15(c_A+s_F)
-\frac{55}{2}\left[c_A\left(\ln\frac{\hat\Lambda}{2}-\frac{36}{11}\ln{\hat m_D(0)}-3.637\right)
\right.
\nonumber\\
&-&\left.\frac{4}{11}s_F\left\{\ln\frac{\hat\Lambda }{2}-2.333
\right\}\right]\hat m_D(0)
+\frac{495}{2}\left[c_A\left(\ln\frac{\hat\Lambda }{2}
+\frac{5}{22}+\gamma\right)
\right.
\nonumber\\
&-&\left.
\frac{4}{11}s_F\left(\ln\frac{\hat\Lambda }{2}-\frac{1}{2}
+\gamma+2\ln2
\right)\right] {\hat m}_D^2(0)
+180 s_F\hat m_q^2(0)\Bigg\} . \label{gap_md0}
\end{eqnarray}
For $m_q$ one obtains
\begin{eqnarray}
 \hat m_q^2(0)=\frac{d_A}{8d_F\left(\pi ^2-6\right)}\frac{\alpha_s s_F}{\pi }\left[3\left(\ln\frac{\hat\Lambda }{2}
+2.198\right)-12\hat m_D(0)\right] . \label{gap_mq0}
\end{eqnarray}
For $\frac{\partial^2}{\partial\mu^2}m_D$ one obtains
\begin{eqnarray}
 && \hspace{-5mm} 45
 \left[2+3\left(\ln\frac{\hat\Lambda }{2}-\frac{7}{2}+\gamma+\frac{\pi^2}{3}\right)
 \hat m_D(0)\right] \hat m_D(0) \hat m_D''(0)
\nonumber\\
&=&\frac{\alpha_s}{\pi } \Bigg\{360 s_F-\frac{55}{2} 
\hat m_D''(0) \left[c_A \left(
\ln\frac{\hat\Lambda }{2}-\frac{36}{11} \ln\hat m_D(0)-6.9097\right)
-\frac{4}{11} s_F \left(\ln\frac{\hat\Lambda }{2}-2.333\right)\right]
\nonumber\\
&& + 495 
\hat m_D''(0) \left[c_A \left(\frac{5}{22}+\gamma+\ln\frac{\hat\Lambda }{2}\right)-\frac{4}{11}
 s_F \left(\ln\frac{\hat\Lambda }{2}-\frac{1}{2}+\gamma+2 \ln2\right)\right]
\hat m_D(0)
\nonumber\\
&& + 20  s_F \left(\ln\frac{\hat\Lambda}
{2}-15.662\right) (24-18 \zeta(3)) \hat m_D(0)
\nonumber\\
&& + 180 s_F \hat m_D(0)^2 \left[7 \zeta(3)+ \left(\frac{\zeta'(-1)}{\zeta(-1)}+\ln\hat m_D(0)+\frac{1}{3}\right) 
(24-18 \zeta(3))\right]
\nonumber\\
&& +360s_F\hat m_q(0)\hat m_q''(0)\Bigg\} . 
\nonumber \\
\label{gap_mdpp0}
\end{eqnarray}
For $\frac{\partial^2}{\partial\mu^2}m_q$ one obtains
\begin{eqnarray}
 \hat m_q(0)\hat m_q''(0)=-\frac{3d_A}{8d_F\left(\pi ^2-6\right)}\frac{\alpha_s s_F}{\pi }\left[44.953+2\hat m_D''(0)\right] .
\label{gap_mqpp0}
\end{eqnarray}
In the expressions above, $m_D(0)\equiv m_D(T,\Lambda,\mu=0)$,  $m_D''(0)\equiv \left.\frac{\partial^2}{\partial\mu^2}m_D(T,\Lambda,\mu) \right|_{\mu=0}$ 
and similarly for $m_q$. 


\section{Quark Number Susceptibility}
\label{qns}

We are now in a position to obtain the second and fourth-order HTLpt QNS following Eq.~(\ref{qns_def}).  
We note that the pure gluonic loops at any order do not contribute to QNS, however, gluons contribute through 
the dynamical fermions through fermionic loops. This makes QNS proportional 
to only quark degrees of freedom.  Below we present \mbox{(semi-)}analytic expressions for both LO and NLO QNS.


\subsection{LO HTLpt second-order QNS}

An analytic expression for the LO HTLpt second-order QNS can be obtained using Eq.~(\ref{pressure-LO})
\begin{eqnarray}
\chi_2^{\rm LO}(T)&=&\left.\frac{\partial^2 }{\partial\mu^2}{\cal P}_{\rm LO}(T,\Lambda,\mu)\right|_{\mu=0}=\frac{1}{(2\pi T)^2}
\left.\frac{\partial^2 }{\partial\hat\mu^2}{\cal P}_{\rm LO}(T,\Lambda,\hat\mu)\right|_{\hat\mu=0}
\nonumber\\
&=&\frac{d_F T^2}{3}\Bigg[1-\frac{3c_F}{4}\left(\frac{g}{\pi}\right)^2
                   +\frac{c_F}{4}\sqrt{3(c_A+s_F)}\left(\frac{g}{\pi }\right)^3                  -
\frac{c_F^2}{64}\left(\pi^2-6\right)\left(\frac{g}{\pi}\right)^4 
\nonumber\\
&&+\ \frac{c_F}{16}(c_A+s_F)\left(\log\frac{\hat\Lambda
}{2}-\frac{7}{2}+\gamma+\frac{\pi ^2}{3}
\right)\left(\frac{g}{\pi }\right)^4\Bigg] \ , \label{chi2_lo}
\end{eqnarray}
where the LO Debye and quark masses listed in Eqs.~(\ref{mass_lo}) and their $\mu$ derivatives have been used. 


\subsection{LO HTLpt fourth-order QNS}

An analytic expression for the LO HTLpt fourth-order QNS can also be obtained using Eq.~(\ref{pressure-LO}) 
\begin{eqnarray}
\chi_4^{\rm LO}(T)&=&\left.\frac{\partial^4 }{\partial\mu^4}{\cal P}_{\rm LO}(T,\Lambda,\mu)\right|_{\mu=0}=\frac{1}{(2\pi T)^4}
\left.\frac{\partial^4 }{\partial\hat\mu^4}{\cal P}_{\rm LO}(T,\Lambda,\hat\mu)\right|_{\hat\mu=0}
\nonumber\\
&=&\frac{2d_F}{\pi^2}\Bigg[1-\frac{3}{4}c_F
\left(\frac{g}{\pi }\right)^2
                +\frac{3}{8}c_F s_F\sqrt{\frac{3}{c_A+s_F}}\left(\frac{g}{\pi
}\right)^3
                -\frac{c_F^2\left(\pi^2-6\right)}{64}\left(\frac{g}{\pi}\right)^4
\nonumber\\
&&              +\ \frac{3}{16}c_F
s_F\left(\log\frac{\hat\Lambda}{2}-\frac{7}{2}+\gamma
              +\frac{\pi^2}{3}\right)\left(\frac{g}{\pi}\right)^4\Bigg] \ , \label{chi4_lo}
\end{eqnarray}
where, once again, the LO Debye and quark masses listed in Eqs.~(\ref{mass_lo}) and their $\mu$ derivatives have been used. We note that both $\chi_2^{\rm LO}$ in (\ref{chi2_lo}) and $\chi_4^{\rm LO}$ in (\ref{chi4_lo}) are the same as those recently obtained by Andersen et al.~\cite{su}; however, the closed-form expressions obtained here have not been explicitly listed therein.


\subsection{NLO HTLpt second-order QNS}
A semi-analytic expression for the NLO  HTLpt second-order QNS can be obtained from Eq.~(\ref{pressure-NLO})
\begin{eqnarray}
 \chi_2^{\rm NLO}(T)&=&\left.\frac{\partial^2 }{\partial\mu^2}{\cal P}_{\rm NLO}(T,\Lambda,\mu)\right|_{\mu=0}=\frac{1}{(2\pi T)^2}
\left.\frac{\partial^2 }{\partial\hat\mu^2}{\cal P}_{\rm NLO}(T,\Lambda,\hat\mu)\right|_{\hat\mu=0}
\nonumber\\
&=& \frac{d_AT^2}{2}\Bigg[\frac{2}{3}\frac{d_F}{d_A} + \frac{\alpha_s}{\pi}s_F\Bigg\{-1
        +4\ \hat m_D(0)
        +\frac{2}{3}\left(\ln{\frac{\hat\Lambda}{2}-15.662}\right)(4-3\zeta(3))\,\hat m_D^2(0)
\nonumber\\ 
      &+& 44.953\ \hat m_q^2(0)
     +\left[\frac{14}{3}\zeta(3)+\left(\frac{\zeta'(-1)}{\zeta(-1)}
      + \ln\hat m_D(0)\right)(16-12\zeta(3))\right]\hat m_D^3(0)\Bigg\}\Bigg] .
      \nonumber \\
\label{chi2_nlo}
\end{eqnarray}
We note that no $\mu$ derivatives of the mass parameters appear in (\ref{chi2_nlo}) and, as a result, $\chi_2^{\rm NLO}(T)$ reduces to such a simple and compact form. This is because the second derivatives of the mass parameters with respect to $\mu$ always appear as multiplicative factors of the gap equations (\ref{gap_md0}) and (\ref{gap_mq0}) and hence these contributions vanish.  Numerically solving for the variational masses using Eq.~(\ref{gap_md0}) and (\ref{gap_mq0}) one can directly compute $\chi_2^{\rm NLO}(T)$ from (\ref{chi2_nlo}).  Alternatively, we have also computed $\chi_2^{\rm NLO}(T)$ by performing numerical differentiation of the pressure in (\ref{pressure-NLO}) which leads to the same result within numerical errors.
 

\subsection{NLO HTLpt fourth-order QNS}

A semi-analytic expression for the NLO HTLpt fourth-order QNS can also be obtained from Eq.~(\ref{pressure-NLO})
\begin{eqnarray}
 \chi_4^{\rm NLO}(T)
                &=&\left.\frac{\partial^4 }{\partial\mu^4}{\cal P}_{\rm NLO}(T,\Lambda,\mu)\right|_{\mu=0}
                 =\frac{1}{(2\pi T)^4}\left.\frac{\partial^4 }
                  {\partial\hat\mu^4}{\cal P}_{\rm NLO}(T,\Lambda,\hat\mu)\right|_{\hat\mu=0}
\nonumber\\
     &=&\frac{d_A}{4\pi^2}\Bigg[ 8\frac{ d_F}{d_A}
     +\frac{\alpha_s}{\pi }s_F 
      \Bigg\{-12 +6  \hat m_D''(0)
\nonumber\\
     &+& 3 \hat m_D^2(0) 
    \left[ \left(\frac{\zeta'(-1)}{\zeta(-1)}+\ln\hat m_D(0)+\frac{1}{3}\right)
     (24-18 \zeta(3)) +7\zeta(3)\right] \hat m_D''(0)
\nonumber\\
    &+& \hat m_D(0)\hat m_D''(0)  \left(\ln\frac{\hat\Lambda} {2}-15.662\right) 
    (8-6 \zeta(3)) 
\nonumber\\
    &-&4 \hat m_D^3(0) \left[31 \zeta(5)-120 \left(
    \frac{\zeta'(-1)}{\zeta(-1)}+\ln\hat m_D(0)\right) (\zeta(5)-\zeta(3))
    \right]
\nonumber\\
    &+& 80\hat m_D^2(0)  
        \left(\ln\frac{\hat\Lambda} {2}-1.0811\right) (\zeta(5)-\zeta(3))
   + 134.859\ \hat m_q(0)\hat m_q''(0)\Bigg\}\Bigg]\, , \;\;
\label{chi4_nlo}
\end{eqnarray}
where the double derivatives of the mass parameters with respect to $\mu$ survive, but the fourth derivatives 
of the mass parameters disappear as discussed earlier. One can now directly 
compute the fourth-order susceptibility by using numerical solutions of the gap equations in (\ref{gap_md0}) 
and (\ref{gap_mqpp0}). Alternatively, we have also computed  $\chi_4^{\rm NLO}(T)$ by performing numerical 
differentiation of the pressure in (\ref{pressure-NLO}) which leads to the same result within numerical errors.


\begin{figure}
 \subfigure{
\includegraphics[width=0.48\textwidth]{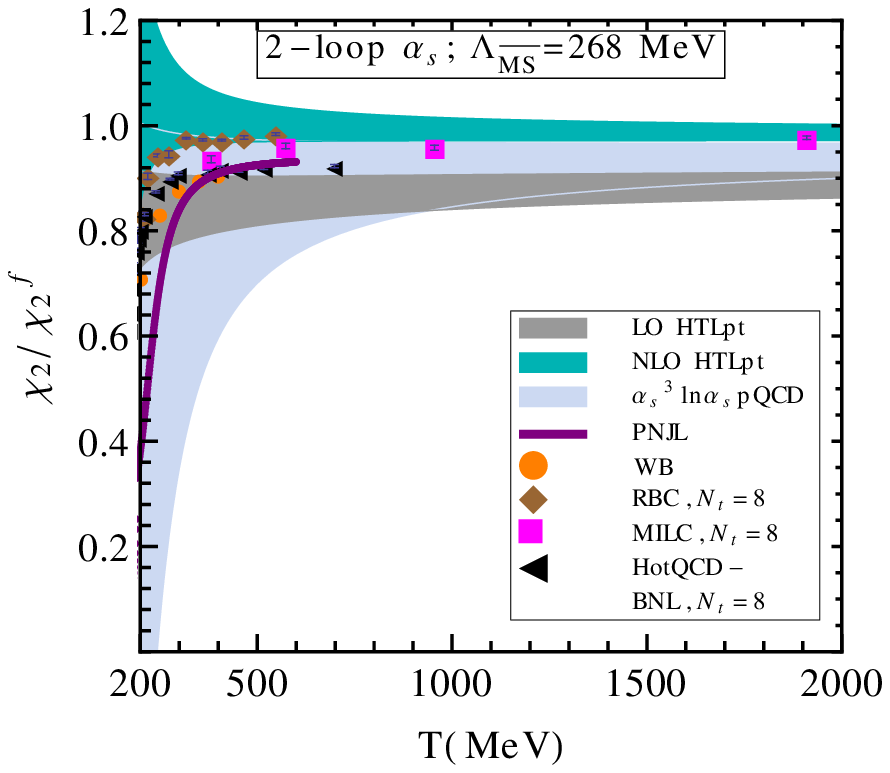}
}
\subfigure{
\includegraphics[width=0.48\textwidth]{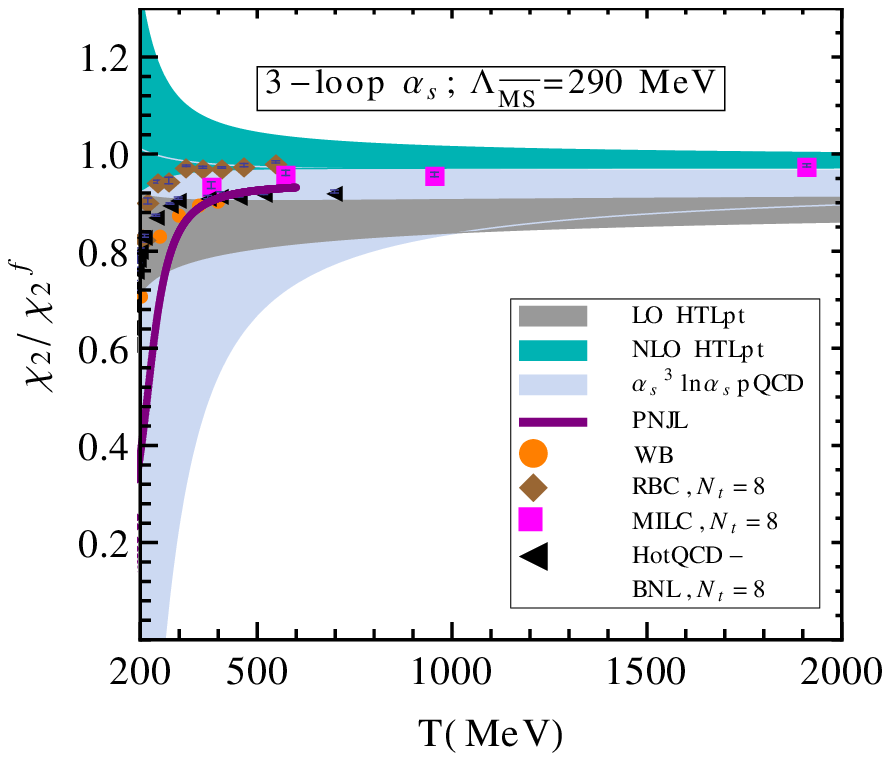}
}
\caption{Left panel: $\chi_2$  scaled by the free field value for LO (grey band) and NLO (sea green band) in 2-loop HTLpt, 
4-loop pQCD (sky blue band)~\cite{vourinen}, LQCD (various symbols)~\cite{wp,peter,peter1,bernard}, and PNJL model 
(thick purple line)~\cite{paramita} are plotted as a function of the temperature.
The bands in HTLpt and pQCD are obtained by varying the ${\overline{\rm MS}}$ renormalisation scale ($\Lambda$) around its 
central value by a factor of two. We also used  $\Lambda_{\overline{\rm MS}}=268$ MeV  and 2-loop $\alpha_s$ for HTLpt and pQCD.
In the PNJL model~\cite{paramita} $\chi_2$ is obtained using a six-fermion interaction. The Wuppertal-Budapest (WB) group~\cite{wp}
data are obtained using the tree-level improved Symanzik action and a stout smeared staggered fermionic
action with light quark masses $\sim 0.035 \, m_s$, with $m_s$ being the strange quark mass near its physical value. The  
RBC-Bielefeld collaboration~\cite{peter} used  a p4 action whereas the MILC collaboration~\cite{bernard} used an asqtad action. 
In both cases the light quark mass ranges from (0.1-0.2)$\,m_s$.
Right panel: Same as left panel but using 3-loop $\alpha_s$ and $\Lambda_{\overline{\rm MS}}=290$ MeV.}
\label{fig_chi2}
\end{figure}


\subsection{Results and Discussions}

Computing the different susceptibilities in HTLpt requires a choice of the renormalization scale $\Lambda$, the $\overline{\rm MS}$ momentum 
scale $\Lambda_{\overline{\rm MS}}$, and a specification of the order of the running coupling $\alpha_s$ used.  In what follows we vary the 
renormalization scale $\Lambda$ by a factor of two around a central value of $\Lambda = 2 \pi  T$ which results in a band 
that can be used to ascertain the level of minimal theoretical uncertainty. The value of $\Lambda_{\overline{\rm MS}}$ depends on the order of 
the running coupling chosen and we fix its value from a recent lattice QCD determination~\cite{latt_lms}.  
The specific value of $\Lambda_{\overline{\rm MS}}$ used in each case is specified in the figure captions.
We show the results obtained using two- and three-loop running; however, one could also use a one-loop running and, after appropriately adjusting $\Lambda_{\overline{MS}}$, the results only show minor differences from the results shown.

In Fig.~\ref{fig_chi2} we have plotted the $N_f=3$ second-order QNS scaled by the corresponding free gas limit as a function of the temperature.
As discussed above, the bands shown for the HTLpt and pQCD~\cite{vourinen} results indicate the sensitivity of $\chi_2$ to the choice of the 
renormalisation scale $\Lambda$. However, $\chi_2$ in both HTLpt and pQCD depends only weakly on the chosen order of the running of the strong 
coupling and in turn only depends weakly on $\Lambda_{\overline{\rm MS}}$, as can be seen clearly from both panels of Fig.~\ref{fig_chi2}.  
The LO HTLpt prediction for $\chi_2$ seems to agree reasonably well with the available Wuppertal-Budapest LQCD data;\footnote{Our result 
in this case is exactly the same as that obtained recently by Andersen et al \cite{su}.} however, there is a sizable variation among different 
lattice computations~\cite{wp,peter1,bernard} considering improved lattice actions and a range of quark masses (see caption).  However, 
lowering the quark mass ($\sim 0.035m_s$, $m_s$ is the strange quark mass) nearer to its physical value~\cite{wp} seems to have a very small 
effect in the temperature range, as seen from the LQCD data.  
Note that for the Wuppertal-Budapest (WB) lattice data shown in Fig.~\ref{fig_chi2}, 
Ref.~\cite{wp} provided a parameterization of their $\chi_2$ data
\begin{eqnarray}
 \chi_2(T)=e^{-(h_3/t+h_4/t^2)} \,f_3 \, [\tanh(f_4 \, t+f_5)+1] \, ,
\end{eqnarray}
where $t=T/(200\;{\rm MeV})$, $h_3 = -0.5022$, $h_4 = 0.5950$, $f_3 = 0.1359$,
$f_4 = 6.3290$, and $f_5 = -4.8303$. The authors of Ref.~\cite{wp} performed the fit for data \cite{wpd}
in the temperature range $125$ MeV $ < T \le$ 400 MeV.  Using the parameterization above, we display their data up to $400$ MeV with a step size of $50$ MeV.  The results for $\chi_2$ obtained using a nonperturbative PNJL model~\cite{paramita} which includes an six-quark interaction are only available very close to the phase transition temperature.

We see in Fig.~\ref{fig_chi2} that NLO HTLpt (\ref{chi2_nlo}) exhibits a modest 
improvement over the pQCD calculation shown, which is accurate to 
${\cal O}(\alpha_s^3\ln \alpha_s)$. However, the NLO $\chi_2$ is higher than the LO one at higher temperature and it goes beyond 
the free gas value at lower temperatures.
It should be mentioned that, although the 2-loop calculation improves upon the LO results by rectifying over-counting which causes incorrect coefficients in the weak coupling limit, it does so by pushing the problem to higher order in $g$.
The reason can be understood in the following way: in HTLpt the loop and 
coupling expansion are not symmetrical, therefore at a given loop order there are contributions from higher orders in coupling. 
Since 
the NLO HTL pressure and thus QNS is only strictly accurate to order ${\cal O}(g^3)$ there is over-counting occurring at higher orders 
in $g$, namely at ${\cal O}(g^4)$ and ${\cal O}(g^5)$. A next-to-next-to-leading order (NNLO) HTLpt calculation would fix the problem 
through ${\cal O}(g^5)$ thereby guaranteeing that, when expanded in a strict power series in $g$, the HTLpt result would reproduce the 
perturbative result order-by-order through ${\cal O}(g^5)$.  

\begin{figure}
 \subfigure{
\includegraphics[width=0.48\textwidth]{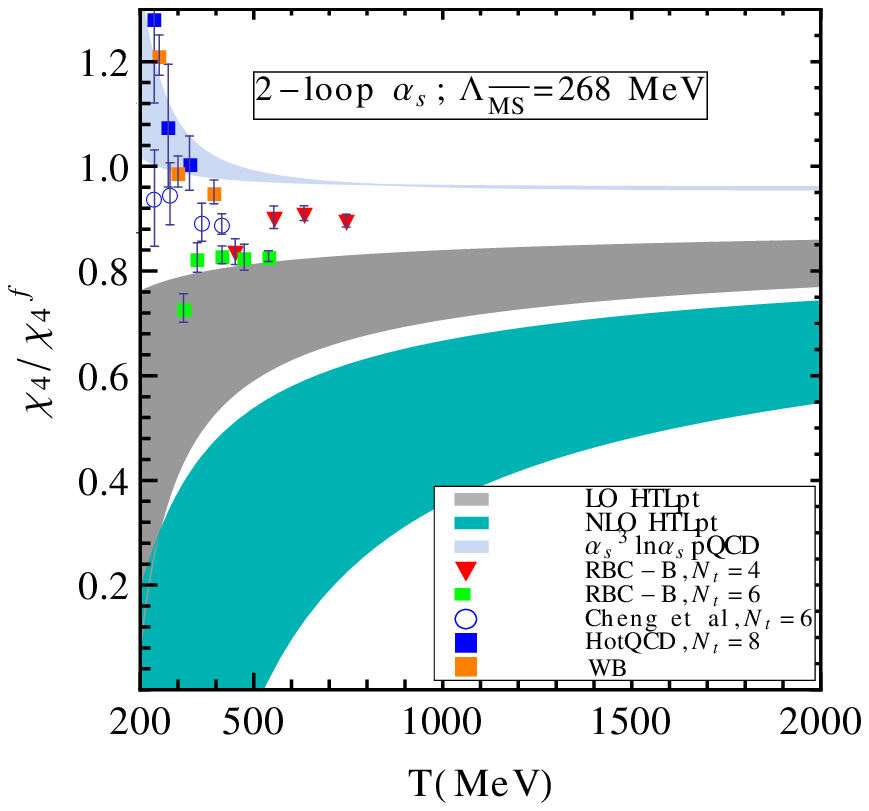}
}
\subfigure{
\includegraphics[width=0.48\textwidth]{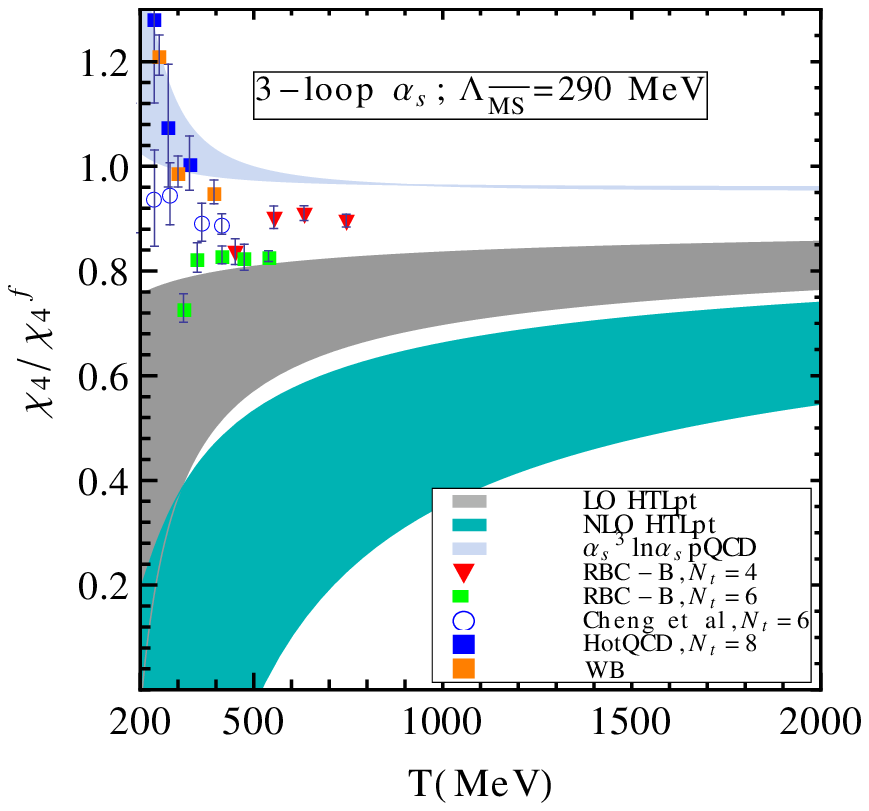}
}
\caption{Left panel: $\chi_4$ scaled by the free field value for LO and NLO HTLpt as given, respectively, in (\ref{chi4_lo}) and (\ref{chi4_nlo}), 
4-loop pQCD~\cite{vourinen}, and
LQCD~\cite{peter1,cheng} are plotted as a function of the temperature. The bands in HTLpt and pQCD are obtained by varying the 
${\overline{\rm MS}}$ renormalisation scale ($\Lambda$) around its central value by a factor of two.  We used  
$\Lambda_{\overline{\rm MS}}=268$ MeV  and 2-loop $\alpha_s$ for HTLpt and pQCD.
Lattice QCD results~\cite{peter1,cheng} are represented by symbols.  The Wuppertal-Budapest (WB) lattice data are taken from Ref.~\cite{borsanyi}.
Right panel: Same as left panel but using 3-loop $\alpha_s$ and $\Lambda_{\overline{\rm MS}}=290$ MeV.}
\label{fig_chi4}
\end{figure}

In Fig.~\ref{fig_chi4} we plot the fourth-order QNS ($\chi_4$) scaled by the corresponding free gas value for HTLpt as given in
(\ref{chi4_lo}) and (\ref{chi4_nlo}), pQCD,  and LQCD.  
Both the HTLpt and pQCD results exhibit a very weak dependence on the choice of order of the running of $\alpha_s$ and thus $\Lambda_{\overline{\rm MS}}$. 
Nevertheless, the HTLpt results are found to be far below the pQCD result~\cite{vourinen} which is accurate to ${\cal O}(\alpha_s^3\ln(\alpha_s))$ and 
the LQCD results~\cite{peter1,cheng}.
Also, the correction to $\chi_4$ when going from LO to NLO is quite large.  This is due to the fact that the fourth order susceptibility is highly sensitive to the erroneous ${\cal O}(g^4)$ and ${\cal O}(g^5)$ terms which appear at NLO.  It is expected that carrying the HTLpt calculation to NNLO would improve this situation; however, only explicit calculation can prove this.  We note additionally that although the pQCD result is very close to the Stefan-Boltzmann limit, the dimensional-reduction resummation method yields a fourth-order QNS which is approximately 20\% below the Stefan-Boltzmann limit \cite{su} which places it slightly higher than the LO HTLpt result shown in Fig.~\ref{fig_chi4}.


\section{Conclusion and Outlook}
\label{concl}

In this paper we have obtained the second- and fourth-order QNS from the NLO HTLpt pressure obtained in a high temperature expansion through 
${\cal O}[(\mu/T)^4]$.  Analytic expressions were found for both $\chi_2^{\rm LO}$ and $\chi_4^{\rm LO}$ within LO HTLpt.  Our result for  
$\chi_2^{\rm LO}$ (\ref{chi2_lo}) is in agreement with the results obtained previously by Andersen et al. \cite{su}.  The LO result 
for $\chi_2$ shows reasonable agreement with available LQCD data~\cite{wp,peter1,bernard}; however, at this point in time there is still a fairly sizable 
variation of this quantity between the different 
lattice groups.  Moving forward it would seem that a detailed analysis of the uncertainties in the various LQCD calculations is necessary before 
detailed conclusions can be drawn.

At NLO we obtained semi-analytic expressions for $\chi_2^{\rm NLO}$ and $\chi_4^{\rm NLO}$ and, after numerically solving the necessary variational 
gap equations for the mass parameters $m_D$ and $m_q$, we obtained our results for $\chi_2^{\rm NLO}$ in (\ref{chi2_nlo}) and 
$\chi_4^{\rm NLO}$ in (\ref{chi4_nlo}).  Unlike the LO results, our NLO calculation takes into account dynamical quark contributions by including 
two-loop graphs which involve fermion loops; however, they suffer from the same problem that NLO HTLpt calculations at zero chemical potential faced: 
the NLO $\chi_2$ in (\ref{chi2_nlo}) gets ${\cal O}(g^3)$ correct but the ${\cal O}(g^4)$ and ${\cal O}(g^5)$ contributions are incorrect if they are expanded 
out in a strict power series in $g$.  As a result, our NLO result for $\chi_2$ scaled to the free limit is closer to unity than the corresponding LO 
result and only shows a weak dependence on the chosen value of the renormalization scale.  Our NLO result for $\chi_4$ (cf, eq.(\ref{chi4_nlo})) in 
which $\mu$ derivatives of the variational mass parameters survive is significantly below the pQCD and lattice data.  

As was the case with the pressure at zero chemical potential, it seems that fixing this problem will require going to NNLO.  In the case of the zero chemical potential 
pressure, performing such a calculation resulted in much improved agreement between HTLpt and LQCD calculations above $\sim 2\,T_c$.  At the very
least a NNLO calculation will fix the over-counting problems through ${\cal O}(g^5)$.
Whether going to 
NNLO will improve the agreement of the HTLpt $\chi_2$ and $\chi_4$ predictions with LQCD results will have to remain an open question for the time being.  
Work on the NNLO calculation has begun, but being a NNLO calculation, care and patience must be applied in equal measure.
  

\acknowledgments
 
We thank Anirban Lahiri for providing the QNS data from the PNJL model, Peter Petreczky and Szabolcs Borsanyi for providing the LQCD data for the QNS, and Nan Su 
for useful discussions during the course of this work.  M.S. was supported by NSF grant No. PHY-1068765.


\end{document}